\documentclass[11pt,epsf,letterpaper]{article}%
\usepackage{color}
\usepackage{amsmath}
\usepackage{amsfonts}
\usepackage{verbatim}
\usepackage{amssymb}
\usepackage{graphicx}%
\providecommand{\U}[1]{\protect\rule{.1in}{.1in}}
\begin{document}

\title{On infrared problems of effective Lagrangians of massive spin 2 fields coupled
to gauge fields}
\author{Fabrizio Canfora$^{1}$ Alex Giacomini$^{2}$, Alfonso R. Zerwekh$^{3}$\\$^{1}$\textit{Centro de Estudios Cient\'{\i}ficos (CECs), Casilla 1469,
Valdivia, Chile.}\\$^{2}$\textit{Instituto de Ciencias F\'{\i}sicas y Matem\'aticas Unversidad
Austral de Chile, Valdivia, Chile.}\\$^{3}$\textit{Departamento de F\'{\i}sica and Centro
Cient\'{\i}fico-Tecnol\'{o}gico de Valpara\'{\i}so } \\\textit{ Universidad T\'ecnica Federico Santa Mar\'ia, Casilla 110-V,
Valpara\'{\i}so, Chile}\\{\small canfora@cecs.cl, alexgiacomini@uach, alfonso.zerwekh@usm.cl}}
\date{}
\maketitle

\begin{abstract}
In this paper we analyze the interactions of a massive spin-2 particles
charged under both Abelian and non-Abelian group using the Porrati-Rahman
Lagrangian. This theory is valid up to an intrinsic cutoff scale.
Phenomenologically a theory valid up to a cutoff scale is sensible as all
known higher spin particles are non-fundamental and it is shown that indeed
this action can be used to estimate some relevant cross section. Such action
necessarily includes St$\overset{..}{u}$ckelberg field and therefore it is
necessary to fix the corresponding gauge symmetry. We show that this theory,
when the St$\overset{..}{u}$ckelberg symmetry is gauge-fixed, possesses a
non-trivial infrared problem. A gauge fixing ambiguity arises which is akin to
the Gribov problem in QCD in the Abelian case as well. In some cases (such as
when the space-time is the four-dimensional torus) the vacuum copies can be
found analytically. A similar phenomenon also appears in the case of Proca
fields. A very interesting feature of these copies is that they arise only for
"large enough" gauge potentials. This opens the possibility to avoid the
appearance of such gauge fixing ambiguities by using a Gribov-Zwanziger like approach.

\end{abstract}

\section{Introduction}

The Standard Model has proven to be much more successful than originally
expected. The 125 GeV boson recently observed by ATLAS and CMS at the LHC
looks very much like the long time awaited Higgs boson. Nevertheless, this
amazing success is today one of the most intriguing puzzles in particle
physics. The resolution of the well known hierarchy and naturalness problems
requires the existence of New Physics at a scale of a few TeV where new
particles must appear. In general, signals for new spin~$0$, $1/2$ and $1$
states have been extensively studied. Nevertheless, particles with higher
spins may also appear. In particular, new massive spin-2 particles are of
phenomenological interest. A well known example is the Kaluza-Klein excitation
of the graviton, predicted in models with extra dimensions. Less attention has
been put on spin-2 particles that can appears as composite states formed by a
pair of color-octet spin-1 fields (colorons) predicted in models like
Top-Color, non-minimal Technicolor and Universal Extra Dimensions
(\cite{Kahawala:2011pc}). Interestingly, in this last case, the massive spin-2
states may be colored. Additionally, in a more standard sector, one of the
most interesting features of strong interactions and QCD is the existence of
many massive higher spin resonances (such as like $\pi_{2}$(1670), $\rho_{3}%
$(1690) or $a_{4}$(2040)) which have a very important phenomenological
role.\newline

Local gauge invariance, which is one of the basic ingredients of the standard
model gives a natural way to couple matter fields to a gauge field. For
instance, in Quantum Electrodynamics, the electron is coupled to the photon by
replacing the partial derivative in the Dirac equation by a gauge covariant
derivative. Naively one may expect that the rule of minimal coupling holds
also for any higher spin field. However it has already been noticed long time
ago by Fierz and Pauli \cite{fierzpauli} that this is not true. Indeed, by
naively replacing the derivative with the covariant derivative in the
equations of motion of any field with spin higher than one immediately gets an
algebraic inconsistency with the equations of motion. To avoid such
inconsistencies several attempts have been done to derive the equations of
motion and the subsidiary conditions for arbitrary spin fields coupled to a
gauge field from an action principle. The only way known up to now to perform
this is by introducing auxiliary fields in the Lagrangian \cite{fierzpauli},
\cite{fronsdal1},\cite{chang}. An explicit Lagrangian for the generic Bosonic
case was proposed in \cite{hagen}.\newline

Unfortunately, also the introduction of auxiliary fields in the Lagrangian
does not leave the theory free of other severe pathologies \cite{aragone},
\cite{velo1}, \cite{velo2}. One well known pathology is that if one insists on
the minimal coupling the theory does not propagate the correct number of
degrees of freedom. \newline

One may be tempted to cure this pathology by introducing non-minimal
couplings. Indeed there have been proposed phenomenological models for spin
two field with spin-stress-energy tensor coupling \cite{suzuki}
\cite{Frank:2012wh} \cite{Geng:2012hy} \cite{Ellis:2012mj}. Another
non-minimal coupling used in phenomenological models is a coupling quadratic
in the spin field and linear in the field strength tensor \cite{Urbano:2012tx}%
. Such a coupling is known as the Federbush model \cite{federbush}. The
introduction of such non-minimal couplings however introduces a new pathology
known as the Velo-Zwanziger problem i.e. super-luminal propagation of the
fields and therefore acausal behavior \cite{velo1}, \cite{velo2}. \newline

On the theoretical side, higher spin particles (whose masses are of the order
of $M_{string}$) have a fundamental role in string theory. In the context of
string theory an interesting possibility to avoid the Velo Zwanziger problem,
in a constant electromagnetic background, has been outlined in \cite{nappi1}
\cite{porrati2}. \newline On phenomenological side, all the experimentally
observed higher spin particles are resonances rather than fundamental
particles: consequently any local Lagrangian describing them is only valid up
to some finite UV cutoff.\newline

Porrati and Rahman \cite{porrati1} have shown that the Velo-Zwanziger problem
can be associated to the existence of an intrinsic cutoff scale in the
Lagrangian describing an interacting massive spin two particle. The authors
analyzed the nature of the UV cutoff and showed that it is possible to
construct an intrinsic, model independent UV cutoff. Their results are based
on the use of the St$\overset{..}{u}$ckelberg formalism since the
St$\overset{..}{u}$ckelberg fields allow to construct gauge-invariant
interactions for charge massive spin-2 fields. It is important to stress that
the St\"{u}ckelberg formalism does not cure by itself the usual pathologies of
interacting higher spin Lagrangians as they are related to the existence of an
intrinsic cutoff. In order to cure these pathologies it is necessary to
introduce non-minimal couplings and additional degrees of freedom. This has
been done for example in \cite{porrati2} \cite{porrati4} for Lagrangians
derived from String Theory. In \cite{porrati1} the cutoff scale can be pushed
to a higher value by adding a new non-minimal coupling in form of a dipole
term. \newline

However in this paper we will show that the use of the Stuckelberg formalism,
introduces also a new problem associated to the gauge fixing. In principle,
one could always gauge-fix the St$\overset{..}{u}$ckelberg fields to zero
using the St$\overset{..}{u}$ckelberg gauge symmetry. However, one of the key
technical points \cite{porrati1} is that, if one chooses \textit{a suitable
covariant gauge fixing}, it becomes possible to diagonalize the kinetic terms
and to single out the sub-sector of the theory which is the source of all
pathologies mentioned above. In other words, this unified description of
phenomena such as strong coupling at finite energy scale, acausal propagation
in external fields and so on within the St$\overset{..}{u}$ckelberg formalism
strictly relies on a specific covariant gauge-fixing\footnote{The
St$\overset{..}{u}$ckelberg formalism also naturally arises when one analyzes
the coupling of spin-2 excitations to gauge field via a Kaluza-Klein reduction
\cite{nappi2}: in this context the so-called St$\overset{..}{u}$ckelberg
symmetry corresponds to the usual symmetry of a spin-2 particle in 5
dimensions when the theory is reduced to 4 dimensions.}.

One of the goals of this paper is to show that this apparently harmless
procedure to choose a covariant gauge-fixing actually hides a non-trivial IR
problem. We will focus on the spin-2 case but our analysis suggests that
similar results hold for spin higher than two. The free part of the
Lagrangians describing the effective theory of a spin-2 charged massive
particle are well known (see, for instance, \cite{hagen} \cite{francia1}
\cite{fronsdal} \cite{porrati3}). The coupling with gauge fields can be
introduced as in \cite{nappi1} and \cite{porrati1}. The non-trivial IR problem
which will be discussed here is related to the pathologies arising when the
St$\overset{..}{u}$ckelberg symmetry is gauge-fixed. It will be shown that the
usual gauge fixing of the St$\overset{..}{u}$ckelberg symmetry is ambiguous.
Such pathologies correspond to gauge-fixing ambiguities of Gribov type
\cite{Gri78} and affect the IR region of the theory. In the present
manuscript, we will mainly analyze the ambiguities arising in the Abelian case
(see, for instance, \cite{kelnhofer}\ and references therein). Interestingly
enough, the way in which such Gribov-like "disease" appears in the present
case suggests a very natural therapy as well. \newline

The paper is organized as follows. First it is shown how the Porrati Rahman
action can be used to estimate some phenomenologically relevant process
involving non-fundamental massive spin particles. Then a basic review of gauge
fixing problems in non-abelian gauge theory will be given. In the fourth
section the gauge fixing problem specific to the Porrati Rahman action and the
presence of Gribov copies will be discussed. The last section will be
dedicated to the conclusions.

\section{Effective Lagrangian for charged spin-2 particles}

\label{sec:pheno}

The standard Pauli-Fierz Lagrangian $L_{PF}$ for a spin-2 massive field on
flat space-times reads:%
\begin{align}
L_{PF}  &  =-\frac{1}{2}\left(  \partial_{\mu}h_{\nu\rho}\right)  ^{2}+\left(
\partial_{\mu}h^{\mu\nu}\right)  ^{2}+\frac{1}{2}\left(  \partial_{\mu
}h\right)  ^{2}-\left(  \partial_{\mu}h^{\mu\nu}\right)  \left(  \partial
_{\nu}h\right)  -\frac{m^{2}}{2}\left(  h_{\mu\nu}^{2}-h^{2}\right)
\ ,\label{PF1}\\
h  &  =h_{\ \mu}^{\mu}\ . \label{PF2}%
\end{align}
The St$\overset{..}{u}$ckelberg procedure corresponds, as explained in
\cite{porrati1},to the replacement
\begin{equation}
h_{\mu\nu}\rightarrow\widehat{h}_{\mu\nu}=h_{\mu\nu}+\frac{1}{m}\left[
\partial_{\mu}\left(  B_{\nu}-\frac{1}{2m}\partial_{\nu}\phi\right)
+\partial_{\nu}\left(  B_{\mu}-\frac{1}{2m}\partial_{\mu}\phi\right)  \right]
\ , \label{stu1}%
\end{equation}
where $B_{\nu}$ and $\phi$ are the so-called St$\overset{..}{u}$ckelberg
fields accounting for the spin-1 and spin-0 degrees of freedom avoiding the
well known singularities of \cite{vandam}. With the above replacement, the
Lagrangian becomes invariant under the following St$\overset{..}{u}$ckelberg
symmetry%
\begin{align}
\delta h_{\mu\nu}  &  =\partial_{\mu}\lambda_{\nu}+\partial_{\nu}\lambda_{\mu
}\ ,\label{stu2}\\
\delta B_{\mu}  &  =\partial_{\mu}\lambda-m\lambda_{\mu}\ ,\label{stu3}\\
\delta\phi &  =2m\lambda\ , \label{stu4}%
\end{align}
where $\lambda$\ and $\lambda_{\mu}$\ are "gauge" parameters.\newline It is
worth to point out that it is also possible to introduce a more generic
Lagrangian as done in \cite{zinoviev} where in Eq. (\ref{stu1}) on the right
hand side there is also a term proportional to $\eta_{\mu\nu}\phi$. This would
add in the gauge transformation (\ref{stu2}) also a term proportional to
$\eta_{\mu\nu}\phi$. We will use however the Lagrangian of \cite{porrati1} due
to its physical interest as it is obtained from a Kaluza-Klein
compactification of the $d+1$ dimensional the free theory. Moreover this
choice does not imply any loss of generality in our discussion as it is
straightforward to see that using the Lagrangian proposed in (\cite{zinoviev})
would not change the results found in this paper (see section 4).\newline

As it is by now well known (see for instance, \cite{porrati1} \cite{nappi1} ),
the use of the St$\overset{..}{u}$ckelberg formalism appears to be unavoidable
if one wants to include the interactions of massive spin-2 particles with
gauge fields which is obviously important for phenomenological studies. Let us
recall that color-octet spin-2 fields are expected to appear as composite
states in many models like Top-color, Technicolor and Universal Extra
Dimensions. Their production at the LHC was studied in \cite{Kahawala:2011pc}
using general properties of bound states. Here, nevertheless, we want to start
by constructing the gauge theory for a color-octet spin-2 massive particle and
examine its properties.\newline

Thus, following \cite{porrati1}, let us consider spin-2 massive particles
charged under a non-Abelian gauge group $h_{\mu\nu}^{a}$ (where $a$ is the
index corresponding to the Lie algebra of the gauge group, we will focus on
$U(1)$ and $SU(N)$ whose structure constants will be denoted as $f^{abc}$). In
this case also the St$\overset{..}{u}$ckelberg fields $B_{\mu}^{a}$ and
$\phi^{a}$\ have to belong to the same representation of $h_{\mu\nu}^{a}$.
Therefore, the St$\overset{..}{u}$ckelberg symmetry becomes%
\begin{align}
\delta h_{\mu\nu}^{a}  &  =\left(  D_{\mu}\lambda_{\nu}\right)  ^{a}+\left(
D_{\nu}\lambda_{\mu}\right)  ^{a}\ ,\label{stu5}\\
\delta B_{\mu}^{a}  &  =\left(  D_{\mu}\lambda\right)  ^{a}-m\lambda_{\mu}%
^{a}\ ,\label{stu6}\\
\delta\phi^{a}  &  =2m\lambda^{a}\ ,\label{stu7}\\
D_{\mu}  &  =\partial_{\mu}+ie\left[  A_{\mu},\right]  \ , \label{stu8}%
\end{align}
where $A_{\mu}$\ is the gauge field to which we want to couple the massive
spin-2 particle.

It is worth emphasizing here two very important differences which distinguish
the gauge parameters of St$\overset{..}{u}$ckelberg symmetry from the usual
gauge parameters appearing in Maxwell and Yang-Mills theories. Firstly, the
St$\overset{..}{u}$ckelberg gauge parameter $\lambda_{\mu}$ carry a Lorentz
index (unlike what happens in Maxwell and Yang-Mills theories). Secondly, the
St$\overset{..}{u}$ckelberg gauge parameters $\lambda$ and $\lambda_{\mu}$ are
also charged\footnote{On the other hand, in the Yang-Mills case, the gauge
parameter only carries indices of the $SU(N)$ group itself while, in the
Maxwell case, the gauge parameter is a real scalar function (and so it carries
no $U(1)$ charge at all).} under the $U(1)$ or $SU(N)$ gauge fields to which
$h_{\mu\nu}$, $B_{\mu}$ and $\phi$ couples. This can be easily seen from Eqs.
(\ref{stu5}), (\ref{stu6}) and (\ref{stu7}). In particular, $\lambda_{\mu}%
^{a}$ has the same structure of $B_{\mu}^{a}$ and, likewise, $\lambda^{a}$ of
$\phi^{a}$. This implies that, in the Abelian case, the St$\overset{..}{u}%
$ckelberg gauge parameters $\lambda_{\mu}$ and $\lambda$ are charged under the
$U(1)$ gauge group (due to the fact that $h_{\mu\nu}$, $B_{\mu}$ and $\phi$
will couple to the Maxwell gauge potential $A_{\mu}$). Consequently, in the
Abelian case, the St$\overset{..}{u}$ckelberg gauge parameters $\lambda_{\mu}$
and $\lambda$ are complex. As it will be shown in the following sections, it
is because of these two differences that, in the case of the St$\overset
{..}{u}$ckelberg gauge transformation, gauge fixing ambiguities appear in the
Abelian case as well.

The replacement corresponding to Eq. (\ref{stu1}) and the minimal coupled
Lagrangian read%
\begin{align}
h_{\mu\nu}^{a}  &  \rightarrow\widehat{h}_{\mu\nu}^{a}=h_{\mu\nu}^{a}+\frac
{1}{m}\left[  \left( D_{\mu}\left(  B_{\nu}-\frac{1}{2m}D_{\nu}\phi\right)
\right) ^{a} +\left( D_{\nu}\left(  B_{\mu}-\frac{1}{2m}D_{\mu}\phi\right)
\right) ^{a} \right]  \ ,\label{stu9}\\
L_{PF}  &  =tr\left[  -\frac{1}{2}\left( \left(  D_{\mu}\widehat{h}_{\nu\rho
}\right) ^{a}\right)  ^{2}+\left( \left(  D^{\mu}\widehat{h}_{\mu\nu}\right)
^{a}\right)  ^{2}+\frac{1}{2}\left( \left(  D^{\mu}\widehat{h}\right)
^{a}\right)  ^{2}+\right. \nonumber\\
&  \left.  -\left(  D^{\mu}\widehat{h}_{\mu\nu}\right) ^{a} \left(  D^{\nu
}\widehat{h}\right) ^{a} -\frac{m^{2}}{2}\left(  \left(  \widehat{h}_{\mu\nu
}^{a}\right)  ^{2}-\left(  \widehat{h}^{a}\right)  ^{2}\right)  \right]
-\frac{Tr}{4}F_{\mu\nu}F^{\mu\nu}\ , \label{stu10}%
\end{align}
where $F_{\mu\nu}F^{\mu\nu}$ is the kinetic term for the gauge field $A_{\mu}%
$. Hence, the Lagrangian in Eq. (\ref{stu10}) describe the effective
interactions of the massive spin-2 particle with a non-Abelian gauge field. As it has been shown in \cite{porrati1} the gauge-fixing terms here below is the best one to study the spectrum of the free theory:

\begin{align}
L_{gf1}  &  =-2\left(  \partial^{\nu}h_{\mu\nu}^{a}-\frac{1}{2}\partial_{\mu
}h^{a}+mB_{\mu}^{a}\right)  \left(  \partial^{\lambda}h_{\sigma\lambda}%
^{a}-\frac{1}{2}\partial_{\sigma}h^{a}+mB_{\sigma}^{a}\right)  g^{\mu\sigma
}\ ,\label{fuffua1}\\
L_{gf2}  &  =-2\left(  \partial^{\nu}B_{\nu}^{a}+\frac{m}{2}\left(
h^{a}-3\phi^{a}\right)  \right)  ^{2}\ , \label{fuffua2}%
\end{align}
then the kinetic terms $L_{free}$ get canonical forms%
\[
L_{free}=h_{\mu\nu}^{a}\left(  \square-m^{2}\right)  h^{a\mu\nu}-\frac{1}%
{2}h^{a}\left(  \square-m^{2}\right)  h^{a}+2B_{\mu}^{a}\left(  \square
-m^{2}\right)  B^{a\mu}+\frac{3}{2}\phi^{a}\left(  \square-m^{2}\right)
\phi^{a}\ .
\]
To describe the interacting case, one needs to replace derivatives with covariant derivatives. Hence, the total Lagrangian can be written as%
\begin{equation}
L=L_{free}-\frac{1}{4}TrF_{\mu\nu}F^{\mu\nu}+L_{int}\ , \label{dipole0}%
\end{equation}
where $L_{int}$ encodes the interactions terms which can be found by simply
expanding explicitly the expressions in Eq. (\ref{stu10}).

In \cite{porrati1} it has been shown that an additional dipole term improves
the intrinsic cutoff scale, moreover it has been shown in \cite{porrati2}
\cite{porrati4} that, if higher powers of the background gauge field are
dropped, an additional dipole term has to be included. Such dipole term in the
Abelian case reads (we follow the notation of \cite{porrati1})%
\begin{align}
L_{dipole} &  =ie\alpha F_{\nu}^{\mu}H_{\mu\rho}^{\ast}H^{\rho\nu
}+c.c.\ ,\label{dipole2}\\
H_{\mu\rho}^{\ast} &  =h_{\mu\rho}^{\ast}+\frac{1}{m}\left(  D_{\mu}B_{\rho
}^{\ast}+D_{\rho}B_{\mu}^{\ast}\right)  -\frac{1}{2m^{2}}\left(  D_{\rho
}D_{\mu}+D_{\mu}D_{\rho}\right)  \phi^{\ast}\ .\label{dipole3}%
\end{align}
The non-Abelian dipole term is the obvious generalization of the above one:
\begin{equation}
L_{dipole} =  \alpha g f^{abc} F_{\nu}^{a\mu}H_{\mu\rho}^{b}H^{c\rho\nu
}
\end{equation}
Although such a term does not affect the analysis of the Gribov problem,
it does affect the computations of physical quantities such as cross sections.
In the following we will give an example of such computations both with and
without the dipole term.

At this point it is convenient to specialize our discussion by assuming that
the gauge group of our theory is the usual color group $SU(3)_{c}$ and
$A_{\mu}$ is nothing else but the gluon. Starting from Lagrangian
(\ref{stu10}), it is, then, possible to obtain the Feynman rules needed for
computing the double production of the color-octet spin-2 particles at the
LHC, at tree level. We focus on double production since, as in any gauge
theory, all the interaction terms contain the matter field (in this case the
spin-2 field) in pairs. In principle, interaction terms coupling, for
instance, two gluons to a single spin-2 particle are possible. Nevertheless,
we recall, such an interaction is not originated by the gauge principle and
introduces an UV and model independent cut-off \cite{porrati1} and theoretical
uncertainties. In this sense, the situation is similar (although more severe)
to the single production of spin-1 color-octet vector resonance which is
plagued of theoretical uncertainties \cite{Zerwekh:2006te}.

\begin{figure}[ptb]
\centering
\includegraphics[scale=0.5]{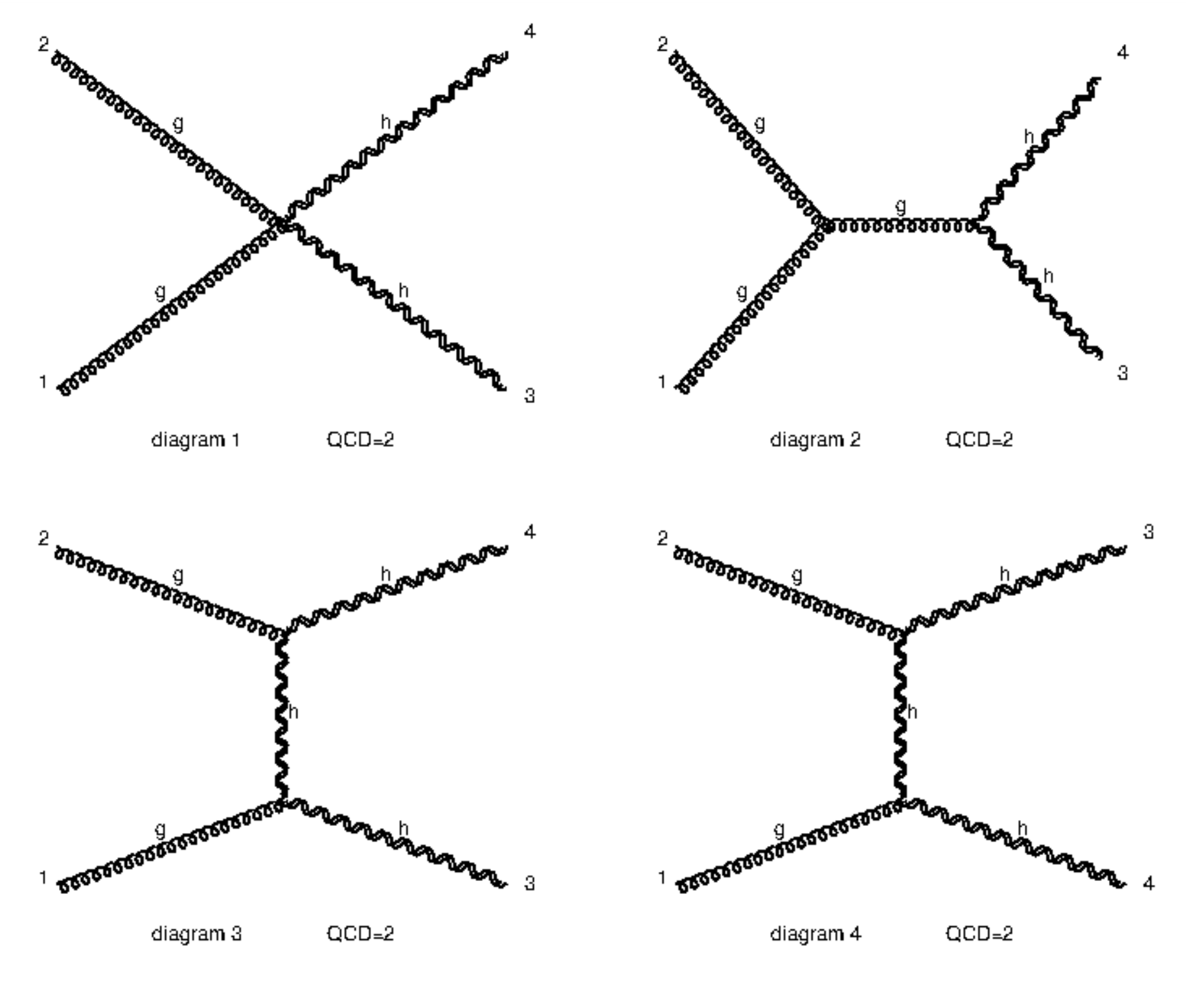} \caption{Feynman diagrams corresponding
to the douple production of the color-octet spin-2 particles ($h$). Figure
produced by Madgraph}%
\label{fig:diagramas}%
\end{figure}

So, we used the package FeynRules \cite{Christensen:2008py} to obtain the
relevant Feynman rules of the model and Madgraph 5 \cite{Alwall:2011uj} in
order to compute the cross section. The Feynman diagrams are shown in figure
\ref{fig:diagramas}

\begin{figure}[ptb]
\centering
\includegraphics[scale=0.4]{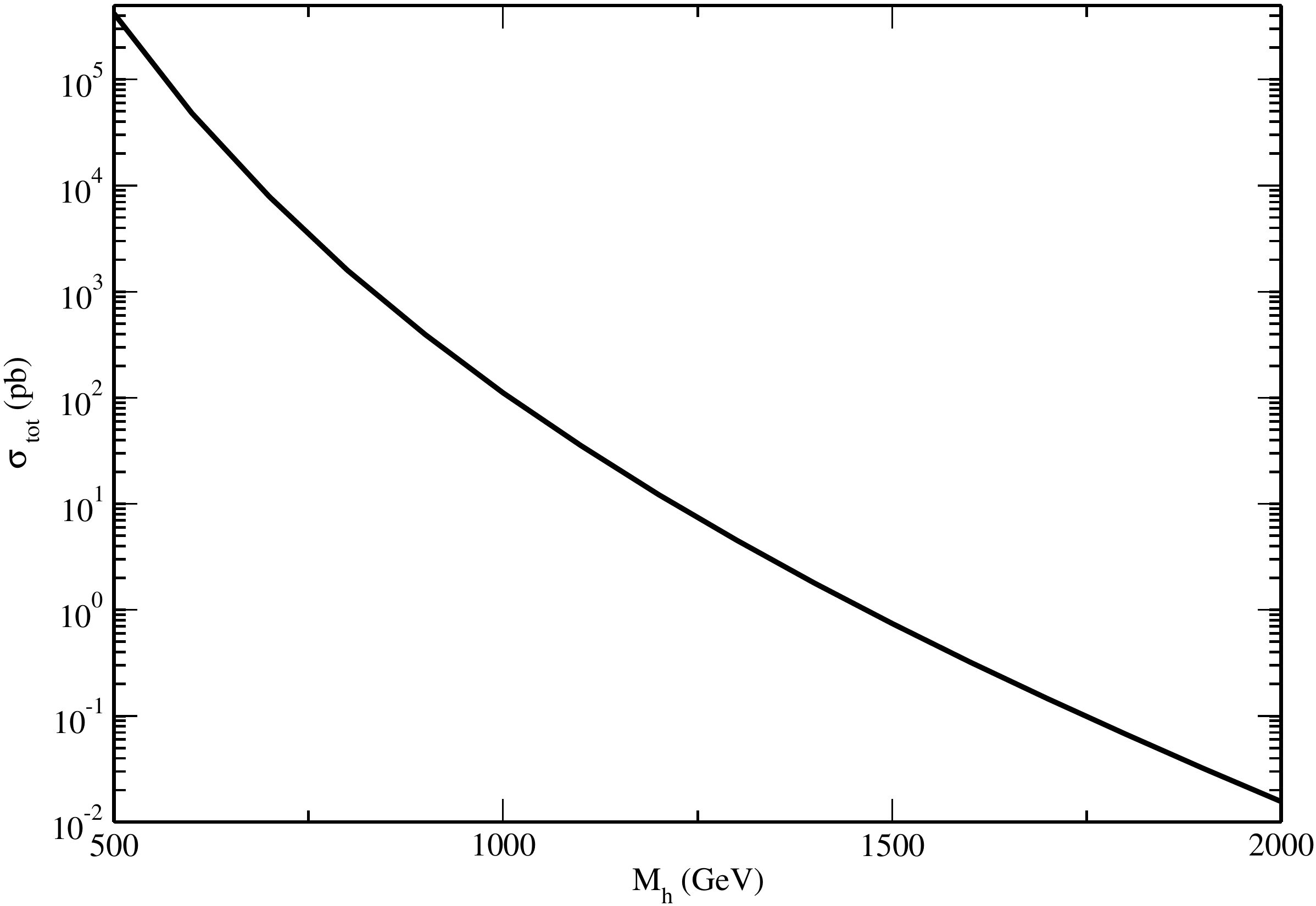} \caption{Cross sections for the double
production of color-octet spin-2 with mass ($M_{h}$) in the range $[500,
2000]$, at the LHC. We assume $\sqrt{s}=14$ TeV and a QCD coupling constant
given by $g_{s}=1.2$}%
\label{fig:xsec}%
\end{figure}

First, we computed the cross sections, without taken into account the dipole term,  for different
masses of the spin-2 particle in the range $[500,2000]$ GeV. The
center-of-mass energy was assume to be $\sqrt{s}=14$ TeV  and we used the CTEQ6L parton distribution function. The obtained cross sections are shown in figure \ref{fig:xsec}.

Our numerical results are in agreement with the ones reported in
\cite{Kahawala:2011pc} which were obtained by a completely different method.

However the inclusion of the dipole term, which is well motivated from the theoretical point of view, may significantly modify these results. In order to evaluate this effect, we computed the cross sections in the context of the LHC as above for different values of the $\alpha$ parameter ($\sigma(\alpha)$) but taking a fixed value for the mass of the  spin-2 particle ($M_h=1$ TeV). In figure \ref{fig:alpha} \label{fig:alpha}, we show the effect of the dipole term by plotting the quantity
\begin{equation}
\frac{\delta \sigma}{\sigma}=\frac{\sigma(\alpha)-\sigma(\alpha=0)}{\sigma(\alpha=0)}
\end{equation}
where $\sigma\alpha)$ is the cross section computed with the dipole term and $\sigma\alpha=0)$ is the cross section without the dipole.

\begin{figure}[ptb]
	\centering
	\includegraphics[scale=0.6]{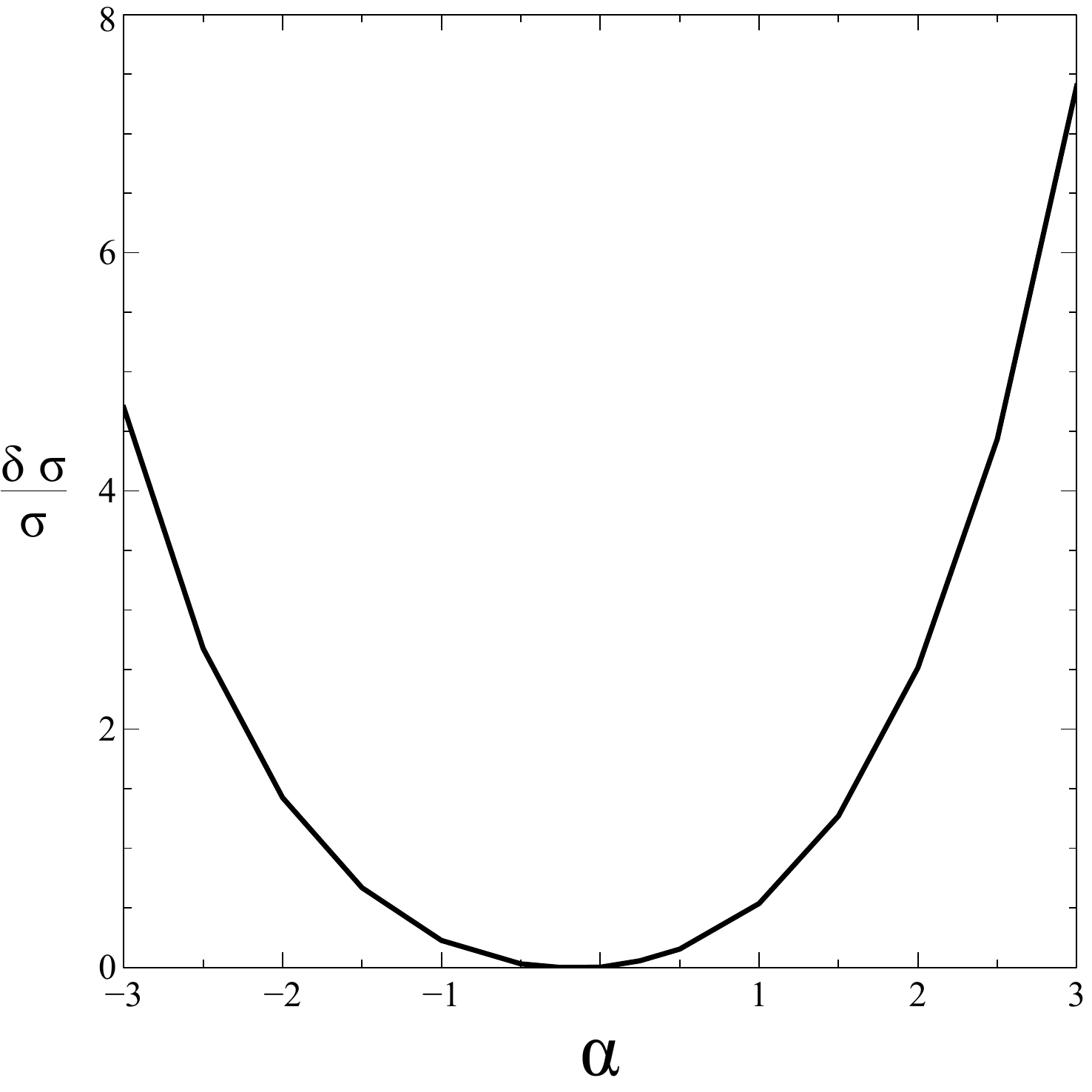} \caption{Effect of the dipole term on the production cross section of two spin-2 particles at the LHC for $M_h=1$ TeV }%
	\label{fig:alpha}%
\end{figure}

Notice that for the preferred value of $\alpha$, at least in the Abelian case, ($\alpha=-1/4$) the departure from the $\alpha=0$ case is neglegible.





In order to proceed, the St$\overset{..}{u}$ckelberg symmetry has to be gauge-fixed.

\section{Review of the SU(N) Gribov problem and its IR nature\ }

The degrees of freedom of any gauge theory are encoded in a Lie algebra valued
one form $\ (A_{\mu})^{a}$. The action functional is invariant under finite
gauge transformations%
\begin{equation}
A_{\mu}\rightarrow U^{-1}A_{\mu}U+U^{-1}\partial_{\mu}U
\label{gaugetranformation}%
\end{equation}
whereas the physical observables are invariant under proper gauge
transformations. Unfortunately (besides the cases of topological field
theories in 2+1 dimensions \cite{wittenjones}), it is still unknown how to use
in practice gauge invariant variable in Yang-Mills case. Hence, the usual
recipe to fix the gauge and to use perturbative expansion around the trivial
vacuum $A_{\mu}=0$ provides one with excellent results when the coupling
constant is small. The most convenient practical choices for the gauge fixing
are the Coulomb gauge and the Landau gauge\footnote{The axial and light-cone
gauge fixings are affected by some non-trivial technical problems in
implementing the physical boundary conditions on the gauge fields: see, for
instance, \cite{DeW03}. Moreover, it is unclear how to carry on the
"$-i\varepsilon$" prescription in the propagators beyond one-loop computations
(see \cite{lieb} for a detailed review).}. However Gribov showed \cite{Gri78}
that a \textit{proper gauge fixing} is not possible globally and that, in the
QCD case, this global effect is very important in the non-perturbative regime.
Furthermore, it has been shown by Singer \cite{singer}, that if Gribov
ambiguities occur in Coulomb gauge, they occur in all the gauge fixing
conditions involving derivatives of the gauge field. In the path integral
formalism, one has to pay close attention to the issue of Gribov copies.
Indeed, as it is well known, there is a close relation between \textit{gauge
fixing ambiguities} and \textit{smooth zero modes of the Faddeev-Popov}
(\textbf{FP}) \textit{operator}. Furthermore, even if one chooses a gauge free
of Gribov copies, the effects of Gribov ambiguities in other gauges generate a
breaking of the BRST symmetry \cite{Fuj}.

One can also verify that all the \textit{simple-minded hopes} that the Gribov
problem can be solved automatically by the path-integral formalism itself
fail. For instance, naively, one could think that if one performs the path
integral without any restriction, the contributions coming from the
non-trivial copies cancel against each others and one would be left with a sum
in which there is just one term for each gauge orbit. Unfortunately, the
reality is totally different and one is confronted with the so-called
Neuberger $0/0$ problem \cite{neuberger}. Namely, in the most direct
translation of BRST symmetry on the lattice, there is a perfect cancellation
among these gauge copies. Consequently, the expectation value of any gauge
invariant (and thus physical) observable in a lattice BRST formulation will
always be of the indefinite form $0/0$ and so ill-defined and, as it is well
known, the formulation of the continuous theory does not help either.

The arising of Gribov copies can be described as a bifurcation problem. Let
$A_{\mu}$ a gauge potential in the Landau gauge and $\left(  A^{U}\right)
_{\mu}$ a potential gauge-equivalent to $A_{\mu}$. In the case of a perfect
gauge fixing it should happen that the system of equations below%
\begin{align}
\partial_{\mu}A^{\mu}  &  =0\ ,\label{stgf1}\\
\partial_{\mu}\left(  A^{U}\right)  ^{\mu}  &  =\partial_{\mu}\left(
U^{-1}A^{\mu}U+U^{-1}\partial^{\mu}U\right)  =0\ , \label{stgf2}%
\end{align}
has a unique trivial solution $U=\mathbf{1}$ for any $A_{\mu}$ satisfying Eq.
(\ref{stgf1}). In other words, one should hope that there is no smooth
globally defined gauge transformation $U$ satisfying Eq. (\ref{stgf2}). As
Gribov showed \cite{Gri78} this is not true. There are known results in the
theory of bifurcation (in particular, the so-called Krasnosel'skii's\ theorem
\cite{berger}) which provide one with sufficient conditions for the appearance
of Gribov copies. As it will be now shown, such conditions have a nice
physical interpretation. In rough terms, the Krasnosel'skii's\ theorem\ can be
stated as follows: write the $U$ in Eq. (\ref{stgf2}) as Taylor expansion%
\begin{equation}
U=\mathbf{1}+\alpha+R(\alpha) \label{expans}%
\end{equation}
where $\mathbf{1}$\ is the identity of the gauge group and $R(\alpha)$
contains terms of order $\alpha^{2}$ or higher. Replacing the above expansion
in Eq. (\ref{stgf2}) one gets%
\[
\partial_{\mu}\left(  A^{U}\right)  ^{\mu}=\left(  \partial_{\mu}D^{\mu
}\right)  \alpha+T(\alpha)=0\ ,
\]
where the operator $T(\alpha)$ (which encodes the non-linear part of Eq.
(\ref{stgf2})) has the property that%
\[
T(\alpha)\underset{\alpha\rightarrow0}{\rightarrow}0\ .
\]
Under some technical assumptions (which are usually verified in situations
which are physically relevant) the Krasnosel'skii's\ theorem tells that, in
order to understand whether or not non-trivial solutions of Eq. (\ref{stgf2})
appear, it is enough to look at the linear part of the equation. In
particular, if the equation%
\begin{align}
\left(  \partial_{\mu}D^{\mu}\right)  \alpha &  =0\ ,\label{stfp}\\
D_{\mu}  &  =D\left(  A\right)  _{\mu}=\partial_{\mu}+\left[  A_{\mu,\cdot
}\right]  \ ,\nonumber
\end{align}
has a smooth normalizable solution then a non-trivial solution of Eq.
(\ref{stgf2}) will appear. The smooth normalizable solutions of Eq.
(\ref{stfp}) can be called \textit{small Gribov copies}. It is well known that
Eq. (\ref{stfp}) (which is nothing but the equation for the zero-modes of the
Faddeev-Popov operator) will have non-trivial solution if the gauge potential
is ``large enough"\footnote{In this case, ``large enough" means large enough
compared to the first eigenvalue of the four-dimensional Laplacian.} with
respect to the a suitable $L_{2}$ norm \cite{Va92}:%
\begin{equation}
\left\Vert A_{\mu}\right\Vert =\int d^{4}xTrA^{2}\ . \label{normcond1}%
\end{equation}
Such norm induces in a natural way the following norm for the zero modes
$\alpha$ of the Faddeev-Popov operator:%
\begin{equation}
\left\Vert \alpha\right\Vert =\int d^{4}xTr\left(  D_{\mu}\alpha\right)
^{2}\ \label{normcond2}%
\end{equation}

The above considerations clarify why, in the case of $SU(N)$ gauge theories
the Gribov is \textquotedblleft Infra-Red" in nature. The reason is that the
relevant quantity for the appearance of zero modes is the above norm for
$A_{\mu}$. In order for the norm to be large enough, it is not necessary that,
when one expands $A_{\mu}$ in Fourier series, there are many Fourier modes
with high (Euclidean) four-momentum $k_{\mu}$. In other words, $A_{\mu}$ can
have a large enough norm even if it is a very slowly varying function with no
Fourier mode with large $k_{\mu}$. Hence, the Gribov problem in $SU(N)$ gauge
theories is an IR issue (of course, this is not the case in gravity but we
will focus on the Yang-Mills case here). On the other hand, the requirement to
have finite norm in the above sense has been often criticized (see, for an
up-to-date discussion \cite{silvionorm}). In particular, it is possible to
construct gauge fields which have finite energy and/or action (and, therefore,
which should not be discarded) but infinite norm in Eq. (\ref{normcond1}).
This suggests that to impose the finiteness of the norms in Eqs.
(\ref{normcond1}) and (\ref{normcond2}) could be a too severe restriction. In
particular, the most conservative requirements that can imposed to avoid
infinitesimal Gribov copies is to ask that both the norm and the energy of the
gauge potential must be finite. We will come back on this point in the next section.

A very elegant solution of the Gribov problem (see, in particular,
\cite{Gri78} \cite{Zw82} \cite{Zw89} \cite{DZ89}\ \cite{Zwa96} \cite{Va92};
two nice reviews are \cite{SS05} \cite{EPZ04}) has been the restriction of the
path integral to the region $\Omega$ around $A_{\mu}=0$ in which the FP
operator is positive (called Gribov region)%
\begin{equation}
\Omega\overset{def}{=}\left\{  \left.  A_{\mu}\right\vert \ \partial^{\mu
}A_{\mu}=0\ \ and\ \ \det\partial^{\mu}D\left(  A\right)  _{\mu}>0\right\}
\ .\label{grireg}%
\end{equation}
In the case in which the space-time metric is flat and the topology is trivial
this approach coincides with usual perturbation theory when the gauge field
$A_{\mu}$ is close to the origin. At the same time, this framework takes into
account the Infra-Red effects related to the partial \cite{Va92}\ elimination
of the Gribov copies \cite{Zw89} \cite{MaggS} \cite{Gracey}. When one
introduces suitable condensates \cite{SoVae2} \cite{SoVar} \cite{SoVar2}
\cite{SoVar3} \cite{soreprl} the agreement with lattice data is excellent
\cite{DOV} \cite{CucM}. Interestingly enough, within this framework, one can
also solve the old problem of the Casimir energy in the MIT-bag model
\cite{canfo2}. The semiclassical approach \textit{a la Gribov} works very well
at finite temperature as well providing one with a good description of the
phase diagram and of the deconfinement transition with results in good
agreement with lattice data \cite{canfo3} \cite{canfo3.5} \cite{canfo3.75}.

On flat space-times with trivial topology, this possibility is consistent with
the usual perturbative point of view since, in the case of $SU(N)$ gauge
theories, it has been shown that there exist \textit{a neighborhood of}
$A_{\mu}=0$ \textit{in the functional space of the gauge potential} (with
respect to a suitable functional norm \cite{Va92}) which is free of Gribov
copies in the Landau or Coulomb gauge. On the other hand, the pattern of
appearance of Gribov copies strongly depends on the space-time metric and
topology and the situation can becomes much more complicated.(see, in
particular,\cite{CGO} \cite{ACGO} \cite{CGO2} \cite{espos1}\ \cite{salgado}
\cite{canfo1}).

\section{The gauge fixing problems of the Porrati-Rahman action}

The unique gauge fixing choice \cite{porrati1} for the St$\overset{..}{u}%
$ckelberg symmetry which allows to diagonalize properly the kinetic terms (in
such a way to provide a unified description of phenomena such as strong
coupling at finite energy scale, acausal propagation in external fields and so
on) is%
\begin{align}
F_{\mu}^{a}  &  =\partial^{\nu}h_{\mu\nu}^{a}-\frac{1}{2}\partial_{\mu}%
h^{a}+mB_{\mu}^{a}=0\ ,\label{gf1}\\
F^{a}  &  =\partial^{\nu}B_{\nu}^{a}+\frac{m}{2}\left(  h^{a}-3\phi
^{a}\right)  =0\ . \label{gf2}%
\end{align}
As it is easy to see, locally this is a good gauge fixing for the
St$\overset{..}{u}$ckelberg symmetry since, \textit{at a first glance},
neither $F_{\mu}^{a}$ nor $F^{a}$ are invariant under the St$\overset{..}{u}%
$ckelberg gauge transformations in Eqs. (\ref{stu5}), (\ref{stu6}) and
(\ref{stu7}). Indeed, under the St$\overset{..}{u}$ckelberg gauge
transformations the gauge-fixing conditions in Eqs. (\ref{gf1}) and
(\ref{gf2}) change as
\begin{align}
\delta F^{a}  &  =\left( \left(  \partial_{\mu}D^{\mu}-3m^{2}\right)
\lambda\right) ^{a} +m\left( \left(  D^{\mu}-\partial^{\mu}\right)
\lambda_{\mu}\right) ^{a}\ ,\label{gf3}\\
\delta F_{\mu}^{a}  &  =\partial^{\nu}\left(  D_{\mu}\lambda_{\nu}+D_{\nu
}\lambda_{\mu}\right)  ^{a}-\partial_{\mu}\left(  D_{\nu}\lambda^{\nu}\right)
^{a}+m\left(  D_{\mu}\lambda-m\lambda_{\mu}\right) ^{a} \ , \label{gf4}%
\end{align}
and for \textit{generic} $\lambda$ and $\lambda_{\mu}$ one has that%
\[
\delta F^{a}\neq0\ ,\ \ \ \delta F_{\mu}^{a}\neq0\ .
\]
Hence, for obvious reasons, we will call \textit{infinitesimal Gribov copies
}$\left(  \lambda_{\mu}\ ,\ \lambda\right)  $\textit{ corresponding to the
St}$\overset{..}{u}$\textit{ckelberg gauge transformations} the non-trivial
solutions of the system $\delta F^{a}=0$ and $\delta F_{\mu}^{a}=0$. The
system of equations for the appearance of infinitesimal Gribov copies
explicitly reads%

\begin{align}
\left( \left(  \square_{FP}-3m^{2}\right)  \lambda\right) ^{a} + m\left(
\left(  D^{\mu}-\partial^{\mu}\right)  \lambda_{\mu}\right) ^{a}  &
=0\ ,\label{gf5}\\
\left( \square_{FP}\lambda_{\mu}\right) ^{a}-m^{2}\lambda_{\mu}^{a}%
+\partial^{\nu}\left(  D_{\mu}\lambda_{\nu}\right)  ^{a}-\partial^{\mu}\left(
D_{\nu}\lambda^{\nu}\right)  ^{a}+m\left( D_{\mu}\lambda\right) ^{a}  &  =0\ ,
\label{gf6}%
\end{align}%
\[
\square_{FP}=\partial_{\mu}D^{\mu}\ ,\ \ \ \square=\partial_{\mu}\partial
^{\mu}\ ,
\]
where in the following the metric will be assumed to be flat and Euclidean.
Obviously, the existence of non-trivial solutions of the above system which
are \textit{smooth everywhere and satisfy reasonable boundary conditions}
implies that the gauge-fixing procedure is not well-defined. As it has been
already emphasized, unlike the pathology analyzed in \cite{porrati1}, this
Gribov-like ambiguity appears in the IR. Here, we will analyze only the
Abelian case since it already contains all the physical ingredients (however,
the present results can be easily generalized to the non-Abelian
case).\newline

The worst gauge fixing pathology which in principle can arise corresponds to
the situation in which non-trivial solutions appear even when $A_{\mu}=0$ in
Eqs. (\ref{gf5}) and (\ref{gf6}). Indeed, if Eqs. (\ref{gf5}) and (\ref{gf6})
would have smooth normalizable solutions for vanishing potential background,
then this would invalidate any perturbative approach to analyze such theory.
Fortunately in this case it can be shown that when $A_{\mu}=0$ no smooth
normalizable solutions of Eqs. (\ref{gf5}) and (\ref{gf6}) exist. In
particular, when $A_{\mu}=0$, Eq. (\ref{gf5}) becomes%
\begin{equation}
\left(  \square-3m^{2}\right)  \lambda=0\ , \label{gf6.5}%
\end{equation}
which does not admit non-trivial solutions due to the fact that the
eigenvalues of Laplacian operator (with any reasonable boundary
conditions\footnote{In mathematical textbooks, the eigenvalue equation for the
Laplacian is usually written as: $\left(  \triangle+\lambda\right)  u=0$,
where $\triangle=\sum_{i}\partial_{i}^{2}$. With this convention, all the
eigenvalues $\lambda_{i}$ are positive: $0<\lambda_{1}\leq\lambda_{2}\leq...$.
Consequently, the operator $\left(  \square-3m^{2}\right)  $ in Eq.
(\ref{gf6.5}) is invertible and the homogeneous equation has only the trivial
solution.}) has positive eigenvalues. Therefore Eq. (\ref{gf6.5}) implies that
$\lambda=0$. It is worth to point out that also if we use the lagrangian of
\cite{zinoviev} instead of the one of \cite{porrati1} it will not change the
further discussion as we will from now on focus on the sector with $\lambda
=0$. If one replaces $\lambda=0$ into Eq. (\ref{gf6}) then one gets
\[
\left(  \square-m^{2}\right)  \lambda_{\mu}=0\ ,
\]
and, due to the positivity of the spectrum of the Laplacian, the only solution
is $\lambda_{\mu}=0$. Hence, these simple considerations show that in order to
have gauge fixing ambiguity for the St$\overset{..}{u}$ckelberg symmetry
$A_{\mu}$ must deviate enough from $0$ in close analogy with the
Gribov-Zwanziger scenario.

It is worth to emphasize the following important point. In the discussion of
Gribov copies in QCD \textit{there are only two key players}: the dynamical
field (that is, the gauge potential $A_{\mu}$) and the gauge parameter
$\alpha$ (see Eqs. (\ref{stfp}) and (\ref{normcond1})). Indeed, the appearance
of gauge fixing ambiguities of the Yang-Mills gauge symmetry only depends on
the norm in Eq. (\ref{normcond1}) characterizing the Yang-Mills gauge
potential itself. In the case of the gauge fixing for the St$\overset{..}{u}%
$ckelberg symmetry, there are \textit{three} \textit{key players}: the
dynamical fields ($h_{\mu\nu}$, $B_{\mu}$ and $\phi$), the gauge parameters of
the St$\overset{..}{u}$ckelberg symmetry (that is, $\lambda$ and $\lambda
_{\mu}$) \textit{and} $A_{\mu}$. This is a huge difference: the appearance of
gauge fixing ambiguity in the St$\overset{..}{u}$ckelberg case (namely, smooth
normalizable solutions of Eqs. (\ref{gf5}) and (\ref{gf6})) \textit{does not
depend on suitable norms of} $h_{\mu\nu}$, $B_{\mu}$ and/or $\phi$ (as one
would expect on the basis of a naive analogy with the standard case). In fact,
as it will be now shown, the appearance of gauge fixing ambiguity in the
St$\overset{..}{u}$ckelberg case \textit{does depend on the norm} of $A_{\mu}$
which is the third key player in the analysis of the gauge fixing of the
St$\overset{..}{u}$ckelberg symmetry. In other words, whether or not gauge
fixing ambiguities appear in the St$\overset{..}{u}$ckelberg case
\textit{depends on the (norm of a) gauge potential of another gauge symmetry}
(Maxwell in the present case). This shows that the analysis of the Gribov
phenomenon in the St$\overset{..}{u}$ckelberg case is more complicated than in
the usual cases due to the fact that it depends heavily with the interactions
with another (non-St$\overset{..}{u}$ckelberg) gauge field. In the appendix we
will also show that a very similar Gribov-like ambiguity appears in the case
of Proca fields.

For simplicity, one can assume $\lambda_{\mu}$ as orthogonal to $A_{\mu}$ in
such a way that the above system reduces to%
\begin{align}
\left(  \square_{FP}-3m^{2}\right)  \lambda &  =0\ ,\label{gf7}\\
\square_{FP}\lambda_{\mu}-m^{2}\lambda_{\mu}+\partial^{\nu}\left(  D_{\mu
}\lambda_{\nu}\right)  -\partial^{\mu}\left(  D_{\nu}\lambda^{\nu}\right)
+mD_{\mu}\lambda &  =0\ ,\label{gf8}\\
A_{\mu}\lambda^{\mu}  &  =0\ , \label{gf8.25}%
\end{align}
where, as it has been already explained in the previous sections, the gauge
parameter $\lambda$ and $\lambda_{\mu}$ are charged under the Maxwell gauge
symmetry. One can take $\lambda=0$ obtaining the following system of equations%
\begin{equation}
\square_{FP}\lambda_{\mu}-m^{2}\lambda_{\mu}+\partial^{\nu}\left(  D_{\mu
}\lambda_{\nu}\right)  -\partial^{\mu}\left(  D_{\nu}\lambda^{\nu}\right)
=0\ . \label{gf8.5}%
\end{equation}
in which the gauge field plays the role of an external background field.

We will now show that gauge fixing ambiguities may arise when the gauge
potential background $A_{\mu}$ is \textquotedblleft large enough". In
particular, we will focus on the most interesting case in which $A_{\mu}$ is
locally a pure gauge. The interest of this choice is that a pure gauge will
pass even the most severe requirements which are usually imposed to avoid the
Gribov issue (see the discussion in the previous section). In particular, a
pure gauge has finite energy and action. Hence, once it is found that a pure
gauge potential can support non-trivial solutions of Eq. (\ref{gf8.5}), there
is no reasonable physical boundary condition which can justify the omission of
such solution. Obviously, by enlarging the class of possible ans\"{a}tze for
$A_{\mu}$ one would also enlarge the number of different non-trivial solutions
of Eq. (\ref{gf8.5}) but our choice is the one which discloses in the clearest
possible way the origin of the phenomenon.

In order to find such copies it is useful to notice that in the case of
\cite{porrati1}, Eq. (\ref{gf8.5}) reads
\begin{equation}
\partial_{\nu}\left(  ieA^{\nu}\lambda_{\mu}\right)  +\partial^{\nu}\left(
ieA_{\mu}\lambda_{\nu}\right)  =-\square\lambda_{\mu}+m^{2}\lambda_{\mu}\ .
\label{gf9}%
\end{equation}

Let us analyze the $U(1)$ St$\overset{..}{u}$ckelberg theory within a
four-dimensional torus $T^{4}$ (this case corresponds to put the system at
finite temperature and in a finite space volume). Such a situation is not just
of academic interest since these circumstances are achieved, for example, in
relativistic heavy ion collisions as those experimentally studied at RHIC and
the LHC (Pb-Pb mode). The natural boundary conditions for $\lambda_{\mu}$ are
the periodic ones. Let us consider in Eq. (\ref{gf9}) a pure gauge potential
$A_{\mu}=const$ (which obviously satisfies periodic boundary conditions). It
is enough to consider $\lambda_{\mu}$ where only two components are switched
on, for example $\lambda_{1}$ and $\lambda_{2}$ where both functions depend
only on the variables $(x_{1},x_{2})$. Using as ansatz%

\begin{equation}
\lambda_{1}=e^{\alpha x_{1}}e^{\beta x_{2}} \label{ansatzlambda}%
\end{equation}
The orthogonality condition $A_{\mu}\lambda^{\mu}=0$ implies
\begin{equation}
\lambda_{2}=-\frac{A_{1}}{A_{2}}\lambda_{1}=-\frac{A_{1}}{A_{2}}e^{\alpha
x_{1}}e^{\beta x_{2}} \label{ortholambda}%
\end{equation}
Choosing $\lambda_{\mu}$ to be divergence free i.e. $\partial_{\mu}%
\lambda^{\mu}=0$ we get
\begin{equation}
\beta=\frac{A_{2}}{A_{1}}\alpha
\end{equation}
so that eventually we get
\begin{equation}
\lambda_{1}=e^{\alpha x_{1}}e^{\frac{A_{2}}{A_{1}}\alpha x_{2}}%
\;\;\;;\;\;\;\lambda_{2}=-\frac{A_{1}}{A_{2}}e^{\alpha x_{1}}e^{\frac{A_{2}%
}{A_{1}}\alpha x_{2}}\ .
\end{equation}

Taking into account that $A_{\mu}$ must be real, it is then straightforward to
check that (\ref{ansatzlambda}) and (\ref{ortholambda}) are solutions of
(\ref{gf9}) which satisfy the correct boundary conditions (i.e. $\alpha$ and
$\beta$ are purely imaginary) only if
\begin{equation}
e^{2}\left(  A_{1}^{2}+A_{2}^{2}\right)  =e^{2}A_{\mu}A^{\mu}>4m^{2}\ .
\label{necond}%
\end{equation}
Hence, smooth solutions of Eq. (\ref{gf9}) satisfying the periodic boundary
conditions (which represent infinitesimal Gribov copies of the St$\overset
{..}{u}$ckelberg gauge symmetry within $T^{4}$) appear when the gauge
potential is large enough.

Interestingly enough, the condition in Eq. (\ref{necond}) is very similar to
the usual condition determining the appearance of Gribov copies in the
Yang-Mills case (when the product ``coupling constant time gauge potential"
must be large enough to give rise to infinitesimal copies). The present
St$\overset{..}{u}$ckelberg Gribov copies are related \textit{both} to the
presence of the massive spin two field \textit{and} to the fact that the gauge
parameter is charged under another (non-St$\overset{..}{u}$ckelberg) gauge
group. \newline

Since (at least in the rather extreme case we have considered) only
\textquotedblleft large" constant gauge potentials generate infinitesimal
Gribov copies of the St$\overset{..}{u}$ckelberg gauge symmetry an intriguing
possibility arises. One could solve the problem using also in this case a
Gribov-Zwanziger like restriction. However, as it has been already explained,
the implementation of the Gribov-Zwanziger restriction in the present case is
more complicated than in the QCD case. On the other hand, it is interesting to
note that one obvious consequence of such restriction would be the appearance
of non-local propagators for the higher spin massive particles. In this sense,
this would not be a surprise since, within the approach developed in
\cite{francia1}, massive higher spin particles can be described without
auxiliary fields the price to pay being the appearance of non-local
propagators. We hope to analyze in future the relations between the two
approaches.\newline

\section{Conclusions and further comments}

In this paper we analyzed the interactions of a massive spin-2 particles
charged under both Abelian and non-Abelian group using the Porrati-Rahman
Lagrangian. Moreover, a scalar field is needed in 4D due to the presence of an
ambiguity in minimal coupling. It has been shown in \cite{porrati2}
\cite{porrati4}\ that if higher powers of the background gauge field are
dropped, an additional dipole term has to be included. In this way, the
well-known inconsistencies (like acausality and wrong number of propagating
degrees of freedom) are under control provided one uses the action only as an
effective action with a characteristic UV cutoff scale. We have shown that,
besides the well understood UV cut-off (signaling the arising of pathologies
such as the Velo-Zwanziger problem), this Lagrangian has also some non-trivial
IR issues. Their origin is a gauge-fixing ambiguity akin of Gribov copies in
QCD for the St$\overset{..}{u}$ckelberg symmetry in an Abelian background.
This type of ambiguity prevents a global covariant gauge fixing of the
St$\overset{..}{u}$ckelberg symmetry (which is the only gauge-fixing choice
unifying phenomena such as strong coupling at finite energy scale, acausal
propagation in external fields and so on). Explicit examples have been found
when the theory is analyzed within a finite volume. In this case, we have
constructed Gribov copies corresponding to the St$\overset{..}{u}$ckelberg
gauge symmetry supported by Abelian gauge potentials with zero field strength
(but large enough norm). To the best of authors knowledge, this is the first
analysis of the peculiar features of the Gribov problem for\ the
St$\overset{..}{u}$ckelberg gauge symmetry. At least in the case of a constant
gauge potential, gauge fixing ambiguities only arise when the gauge potential
is \textquotedblleft large enough" (as it happens in QCD). From the
phenomenological point of view, this is a fortunate circumstance since it
allows perturbative analysis, like the one presented in section
\ref{sec:pheno}, which can be useful for the experimental search of colored
spin-2 states predicted by some models. From a theoretical perspective, it is
natural to wonder whether the Gribov-Zwanziger strategy of restricting the
path integral to a copy free region can be applied in this case as well. We
hope to come back on this interesting issue in the future.\newline

\section{Acknowledgments}

The authors thank Dario Francia and the anonimous referee for their many
useful comments and suggestions which improved a lot this manuscript. This
work is partially supported by FONDECYT grants 1160137, 1150246 and 1160423, 
Conicyt Proyecto Basal FB0821 and ACT1406.
The Centro de Estudios Cient\'{\i}ficos (CECs) is funded by the Chilean
Government through the Centers of Excellence Base Financing Program of
CONICYT. \appendix

\section{The Proca field}

In this appendix we will show that a similar Gribov-like phenomenon also
appears for Proca fields. The results in the main text together with the
discussion in this appendix suggest that this phenomenon could be relevant for
higher spin fields as well. Here we will follow the presentation of the
charged Proca field outlined in \cite{porrati1}. Let us consider the usual
complex Proca Lagrangian%
\[
L=-\frac{1}{2}G_{\mu\nu}^{\ast}G^{\mu\nu}-m^{2}W_{\mu}^{\ast}W^{\mu
}\ ,\ G_{\mu\nu}=\partial_{\mu}W_{\nu}-\partial_{\nu}W_{\mu}\ .
\]
The action becomes gauge invariant after the replacement%
\[
W_{\mu}\rightarrow V_{\mu}-\frac{\partial_{\mu}\phi}{M}\ .
\]
Then, as in the spin-2 case discussed in the main text, the coupling with a
$U(1)$ field $A_{\mu}$ is achieved introducing the covariant derivatives%
\[
\partial_{\mu}\rightarrow D_{\mu}=\partial_{\mu}\pm ieA_{\mu}\ .
\]
One can obtain a diagonal kinetic term including the gauge-fixing term%
\begin{equation}
L_{gf}=-\left\vert \partial_{\mu}V^{\mu}-M\phi\right\vert ^{2}\ .
\label{proca3}%
\end{equation}
Once again, it is worth emphasizing that unlike what happens in Yang-Mills
theory, in the present case such a gauge-fixing is mandatory in order to have
a well-defined kinetic term.

However, due to the coupling with the $A_{\mu}$ field, the gauge-fixing in Eq.
(\ref{proca3}) does have Gribov copies. Following the same steps as in Eqs.
(\ref{gf1}), (\ref{gf2}), (\ref{gf3}), (\ref{gf4}), (\ref{gf5}) and
(\ref{gf6}) one gets the following \textit{equation for the Gribov copies of
the charged Proca field}:%
\begin{equation}
\partial_{\mu}D^{\mu}\alpha-M^{2}\alpha=0\ ,\label{proca4}%
\end{equation}
where $\alpha$ is the complex $U(1)$ gauge parameter. As in the case analyzed
in the main text (see Eq. (\ref{gf6.5})), non-trivial solutions only appear
when the $U(1)$ gauge field is "large enough" since Eq. (\ref{proca4}) has
only trivial solutions when $A_{\mu}=0$ (and, by continuity, when it is
small). On the other hand, it is easy to constrcut non-trivial solutions
following the same approach outlined in the previous sections in $T^{4}$.
Moreover, in the present case of charged Proca fields, it is also possible to
construct many explicit examples of copies by applying the Henyey\ strategy
\cite{heniey}\ (which cannot be applied so easily in the spin-2 case). Namely,
one can interpret Eq. (\ref{proca4}) as an equation for $A_{\mu}$ choosing, a
priori, the $\alpha$. In this way, one obtains explicit examples of copies
together with the corresponding $U(1)$ gauge fields supporting them. Indeed,
in the case of the Proca field, it is simpler than in the spin 2 case to find
explicit copies as using the Henyey approach one has only two equations to
solve-one for the real and one for the immaginary parts of $\alpha$-and four
components of $A_{\mu}$ at one's disposal to play with.

\end{document}